# Cycles of Discourse, Speech Dysfluency, and Active Inference

Thomas Parr[1*], Birtan Demirel[2], Youssuf Saleh[1], Sanjay Manohar[1,3]

[1] Nuffield Department of Clinical Neurosciences, University of Oxford, UK
[2] Institute of Biomedical Engineering, Department of Engineering Sciences, University of Oxford, UK
[3] Department of Experimental Psychology, University of Oxford, UK

[*]Correspondence: thomas.parr@ndcn.ox.ac.uk

**Abstract**

Speech is a complex motor and social act. We must not only produce and understand sequences of variable-length words formed of ordered syllables but also speak and listen in turn. This theoretical paper introduces a computational model of speech production and auditory segmentation based upon sequences of discrete phonemes. The purpose of this is to develop a vehicle to test (in silico) hypotheses about the mechanisms that govern loss of speech fluency, something that may happen transiently in people who stutter or progressively in various neurodegenerative conditions. The modelling here appeals to Partially Observable Markov Decision Processes that provide a useful generic formulation for specifying the internal model our brains might use to describe dynamic environments in which they can exert partial control and make partial observations. The core features of the specific model used here are: (1) a reciprocal mapping from words to sequences of varying numbers of phonemes (2) a common subset of states for both speech generation and speech perception. A further novel feature is a social component, in which one must infer who is speaking (self or other). We propose a series of hypotheses about the computational mechanisms that might underwrite speech breakdown. For example, we demonstrate that reducing the precision in the next word introduces pauses and start-of-word repetitions of the sort associated with stuttering. We examine the patterns of behaviour and belief-updating predicted by each of these hypotheses, with implications for both speaking and listening. Finally, we discuss plausible neural substrates for the underlying message passing that might be amenable to interrogation with functional imaging.

## 1—Introduction

How do we know, if speaking, when we have finished one word to be able to move to the next? How do we know, if listening, when another speaker has finished a word and moved to the next? The first question is important in the fluent generation of speech while the second matters in segmenting words from sequences of sounds when we are listening. This paper serves as a technical note that proposes a duality between the solutions to these two questions and that has implications for the fluency of speech. It may be seen as complementary to a recent conceptual analysis of stuttering predicated upon active inference, in which uncertainty about timings in perceptual cycles takes centre stage (Usler, 2025). Our intention is that the illustrative model introduced here will be used in subsequent work to examine the possible mechanisms that lead to a breakdown in speech fluency, including stuttering and non-fluent aphasias.

A core problem in both speaking and listening is that there are multiple different timescales in play (Arnal et al., 2016; Donhauser & Baillet, 2020). Phonemes progress faster than words, which progress faster than sentences. While one could try to account for this by appealing to an internal pacemaker (Cannon, 2021; Parr et al., 2025), or to use temporally deep hierarchies (Friston et al., 2017c; Hasson et al., 2008; Heilbron et al., 2022; Kiebel et al., 2008; Pezzulo et al., 2018), the difficulty is that the timings for speech are not regular. For instance, different words include different numbers of phonemes (Fenk-Oczlon & Pilz, 2021; Grotjahn & Altmann, 1993; Marian et al., 2012). This means one cannot naively factorise the states of an internal clock from those of the word they wish to speak or infer they have heard. One must account for the entanglement between timings and words (Burton et al., 2000). For an interesting account of Bayesian solutions to this kind of problem more generally, see (Kaltenmaier et al., 2025).

There is a vast literature on speech production and auditory segmentation. While it would be impossible to do justice to this in the space available here, the following offers a brief overview of some of the important advances upon which this paper builds. Models of speech production have emphasised the importance of speech gestures (Saltzman & Munhall, 1989), which play an analogous role to the dynamic movement primitives (Schöner, 1990) hypothesised in movement neuroscience to construct complex continuous movements from discrete elements. Effectively, gestures (or movement primitives) represent attracting points for the set of articulators and vocal tract (or effector muscles in limbs). By sequentially activating different attractors, one can blend a series of gestures into continuous speech production. The way in which gestures implicitly coordinate the articulators determines the 'utterances' generated (Browman & Goldstein, 1992).

A further interpretation of these speech dynamics (Perrier et al., 1996) inherits from equilibrium point theories of motor control (Feldman & Levin, 2009; Feldman & Orlovsky, 1972), in which the attractors associated with each speech gesture are understood as equilibrium points of a dynamical system, that are shifted with time to generate complex behaviour. The tendency towards an equilibrium point has the appearance of proprioceptive error-correction. Equilibrium point theories have been taken to imply that rest positions in speech are themselves equilibrium points and should therefore be achieved consistently, but that rest articulator positions (or 'articulatory settings') should vary between in different languages that require different shifts in equilibrium point—something that has been demonstrated using x-ray imaging of the articulators at rest (Gick et al., 2004).

While error-correction from auditory feedback appears to be an important mechanism by which we modulate our speech (Mitsuya & Munhall, 2019), it is only part of the picture. The importance of extra-auditory sensory modalities in speech perception and production have been demonstrated through an innovative experimental design in which auditory and somatosensory consequences of speech production were independently manipulated (Ito et al., 2020), demonstrating the importance of somatosensory information in speech learning.

In parallel with a (soft) discretisation of speech generation, into sequences of gestures or equilibrium points, there has been development of theories of 'chunking' or segmentation of heard speech. These theories often depend upon temporal integration windows, the durations of which may vary with different oscillatory brain rhythms (Poeppel, 2003)—with different windows accounting for the capture of different timescales of speech. Interestingly, the temporal windows of particular relevance for tracking speech envelopes appear to be within the theta range (~4-8Hz) (Giraud & Poeppel, 2012), compatible with the timescales proposed for speech gestures (Browman & Goldstein, 1992). Recent single neuron electrophysiology findings in humans reinforce this, with evidence that the activity of neurons in the dominant prefrontal cortex can represent segmented phonetic sequences (Khanna et al., 2024), and that transient phase-locking of subthalamic nucleus to cortical oscillations occurs during syllable repetition with a transient period consistent with the theta range (Vissani et al., 2025).

In short, both speech generation and segmentation appear to rely upon implicit sequences of events—either gestures or temporal integration windows—that occur over comparable timeframes. Both rely upon auditory data (generated by self or other), with speech generation requiring an element of error correction, also involving proprioceptive (and somatosensory) feedback that has been interpreted according to Equilibrium Point frameworks. Our approach in this paper draws from the sequential formulations of speech outlined above and focuses upon a timescale compatible with phonemic sequences in both generation and segmentation of speech.

Conceptually, the model we will present has much in common with motor accounts of speech perception rooted in mirror neuron theories of social interaction and mimicry (D'Ausilio et al., 2009; Fadiga & Craighero, 2006; Liberman & Mattingly, 1985)—at least, with the weaker formulations of such theories (Lotto et al., 2009). Of note, stronger versions of these theories are controversial [see (Cook et al., 2014; Hickok, 2009a) for detailed discussions] and our intention is not to directly support or refute these here. While we are not suggesting that speech perception requires motor machinery, the formulation on offer here implies some of the same neural machinery is in play during both speech production and speech perception [see (Möttönen & Watkins, 2009; Watkins et al., 2003) for evidence, using transcranial magnetic stimulation, that motor regions appear to play a role in speech perception]. It would be just as accurate to say that we treat speech production using a perceptual theory as to say we treat speech perception using a motor theory. Our approach also has much in common with other approaches to speech generation (Guenther, 2016). Specifically, the idea that speech generation relies upon error correction coheres with the active inferential perspective that motor output is a form of prediction error correction (see (Bradshaw et al., 2024) for a comprehensive account of the relationship between these formulations)—i.e., movement is such that it brings proprioceptive data in line with their anticipated values. However, the social element, allowing both speech generation and perception, goes beyond purely motor accounts.

Our treatment of common pathways of speech production and perception is partly motivated by observations that people who stutter often find it easier to speak fluently when they are on their own (Jackson et al., 2021). The implication is that there may be something inherently more difficult about formulating elements of speech in the presence of others. This may be a generic issue in the context of social interaction (Rittershofer et al., 2025). The answer we propose is that, given conversations rely upon a form of turn taking—something which has been extensively investigated in psychology (Corps et al., 2022; Heldner & Edlund, 2010; Levinson, 2025; Levinson & Torreira, 2015; Meyer, 2023; Sacks et al., 1974; Skantze, 2021)—in which one person speaks at a time, the cycles that determine the start and end of a word are shared between the conversational participants (Friederici, 2002). In principle, this means that, in the presence of another, one has the added challenge of determining who sets that cycle—perhaps making sense of the benefits of metronome-paced speech in people who stutter, in that a metronome unambiguously sets the cycle (Fransella & Beech, 1965; Howell & El-Yaniv, 1987; Toyomura et al., 2015). This also means that any computational lesion proposed under any model in which the cycle of speech generation is shared between conversational participants makes predictions about both speech perception and production, providing opportunities to test predictions in multiple domains.

It is worth saying upfront what we will not address. We take a reductionist approach here to render the problems tractable. Specifically, while we aim to capture some of the statistics of speech, we will use a minimal set of rules that generate sequences of words, but that do not deal with syntactic rules or semantics of a meaningful sort. In addition, we note that various levels of description—e.g., syllable level selection and initiation in 'GODIVA' based accounts (Civier et al., 2013) of stuttering depend upon syllable-level motor programs. We instead focus upon discretised phonemes as the basic unit of speech addressed by our modelling. This is not to say that there is something special about phonemes, and we note there is controversy about whether phonemes are meaningful units, or are epiphenomenal, arising from more fundamental units of speech (see above on 'speech gestures'). This level of modelling does not negate the importance of sub-phonemic levels of description. However, it is justified in our setting in that part of our explanatory target is the observed phonemic omissions and repetitions observed in some dysfluency syndromes. Furthermore, it can be motivated in light of physiological observations that encoding of phonemic features can be identified from the superior temporal gyrus in humans (Mesgarani et al., 2014)—whether or not these are built from more fundamental units. While extensions to this modelling to incorporate these are possible, we have kept things deliberately simple to try to draw out core aspects of speech fluency. This simplicity facilitates extensions in other directions, and particularly the social element of speech—whether one is talking to themselves, to another, or listening to another.

In what follows, we first provide a brief conceptual overview of the theoretical framework—Active Inference—we will use. This involves setting out the generic aspects of the generative model architecture and Bayesian belief propagation that we will use to simulate behaviour. We then introduce the specific features of the model on offer in this paper. These include the presence or absence of a conversational partner, which plays a key role in assessing the impact of some social aspects of speech. In addition, our generative model allows for a dynamic interplay between sequences of (pseudo)words and sequences of phonemes within each word that allows one to segment (or generate) words of different lengths. We next demonstrate through illustrative simulations that this same model can be used to generate fluent speech, including words of varying lengths, and to successfully segment heard speech to identify the words being spoken. Finally, we propose several alternative hypotheses as to the mechanisms of breakdown of fluent speech and simulate speech generation and perception under each of these mechanisms. For example, we demonstrate that reducing the precision of predictions about the next word introduces pauses and start-of-word repetitions of the sort associated with stuttering.

## 2—Theoretical background

In what follows, we appeal to simulations based upon Partially Observable Markov Decision Processes (POMDPs) solved using Active Inference (Buckley et al., 2017; Friston et al., 2015). POMDPs have a long history (Åström, 1965; Kaelbling et al., 1998) of application to a range of decision making problems, and form the basis for ideas of planning as inference (Botvinick & Toussaint, 2012). They have been used specifically in models addressing speech (Young et al., 2013) and hearing (Holmes et al., 2021). The details of the generic aspects of POMDPs as solved using Active Inference are unpacked in detail elsewhere (Parr et al., 2022; Sajid et al., 2021a; Smith et al., 2021), but a brief introduction to core concepts (see Table 1 for a glossary) is important in understanding what follows. The central idea is that one can model brain function as if our brains are engaged in inference using an internal world model (Çatal et al., 2020; Diester et al., 2024; Pezzulo et al., 2024). Under this approach, the problem of understanding behaviour is reframed as the problem of finding the form of the brain's world model. Once formulated, any given model can be inverted to simulate belief-updating[1] and behaviour based upon standard inferential message passing schemes (Friston et al., 2017b; Schwöbel et al., 2018; Yedidia et al., 2005)—whose sparse dependency structure share much with the sparse network architectures observed in synaptic communication (Bastos et al., 2012; George & Hawkins, 2009; Parr et al., 2019). Such simulations allow us to examine the predictions a given model makes about behaviour, the internal brain dynamics that realise that behaviour, and the effect of perturbations on the above.

**Table 1—Glossary**

| Term | Definition |
|---|---|
| **Bayesian Inference or Model Inversion** | Bayesian inference is sometimes referred to as a model inversion process, in that it takes prior plausibility of explanatory variables, combined with the likelihood that some data would be observed given these explanations, and inverts the |

[1] Beliefs here refer to Bayesian beliefs or probability distributions—these are not necessarily conscious or declarative beliefs.

| | |
|---|---|
| | probability distributions to give a posterior probability of explanation given data. |
| **(Perceptual) inference** | Perceptual inference, here, is taken to be the use of Bayesian inference to obtain a posterior probability of the causes of one's sensory data. |
| **Active Inference** | This is the process by which one fits a model to sensory data both by optimising one's beliefs (posterior probabilities) about the causes of those data and by acting upon the world to bring sensory data in line with one's predictions. |
| **Internal world model or Generative model** | These terms refer to the prior probabilities (including priors about dynamics over time) and likelihoods (conditional probabilities of data given their causes) that jointly specify a probability distribution capturing the way in our brain's implicitly 'believe' the world generates our sensory data. |
| **Partially Observed Markov Decision Process (POMDP)** | These are a specific class of generative or world models that make the following assumptions: (1) sensory data at each moment in time are generated from hidden (unobservable) states which (2) evolve such that each state depends only upon its temporal predecessor and (3) upon the choice made at the previous point in time. Such models are normally equipped with a value function for preferred (e.g., rewarding) sensory data. Active Inference formulations place priors on choices such that the most likely choices maximise anticipated value and the potential for uncertainty resolution. |
| **Belief-updating** | Posterior probabilities are often referred to as 'beliefs', normally of an implicit or unconscious nature. As new data become available, these beliefs can be further updated via Bayesian inference. |
| **Message passing** | Just as our brains use synaptic communication, by passing messages from one neuron to another, many approaches to Bayesian inference also rest upon forms of message passing in which one passes information between nodes of a network. This is particularly relevant for generative models comprising multiple interacting variables. The message passing scheme used to perform Bayesian inference here is known as belief propagation and is detailed in Figure 1. |
| **Sparse dependency or sparse network** | Sparsity in this context is the property that, in a network, every node is not connected to every other node. This is a property of brain connectivity, but also of POMDP generative models in which some variables are said to be conditionally independent of others. Conditional independence here means that, for two variables in a probabilistic model, it is possible to factorise the |

joint probability distribution such that there is no factor containing both variables.

The form of the models used for Active Inference is typically a probabilistic generative model comprising prior beliefs about hidden, unobservable, states and their dynamics, and conditional probability distributions that take us from those hidden states to the observable sensory data they cause. The dynamics of hidden states are treated as being dependent upon the course of action, or policy, that our simulated conversational participant may select. Policies are more likely to be chosen when they solicit preferred observations or informative observations that facilitate greater certainty about the world (Sajid et al., 2021b). Actions are generated by weighting predictions about observations one step ahead in time based upon the probability of different policies to construct a predictive distribution from which the most likely observation is sampled for those modalities over which the simulated participant has control (Friston et al., 2020)—in keeping with a reflexive form of motor control based upon the fulfilment of proprioceptive predictions (Adams et al., 2013). Figure 1 shows the form of the generic POMDP structure implied by the above and the (belief-propagation) update scheme for each factor used in our simulations here.

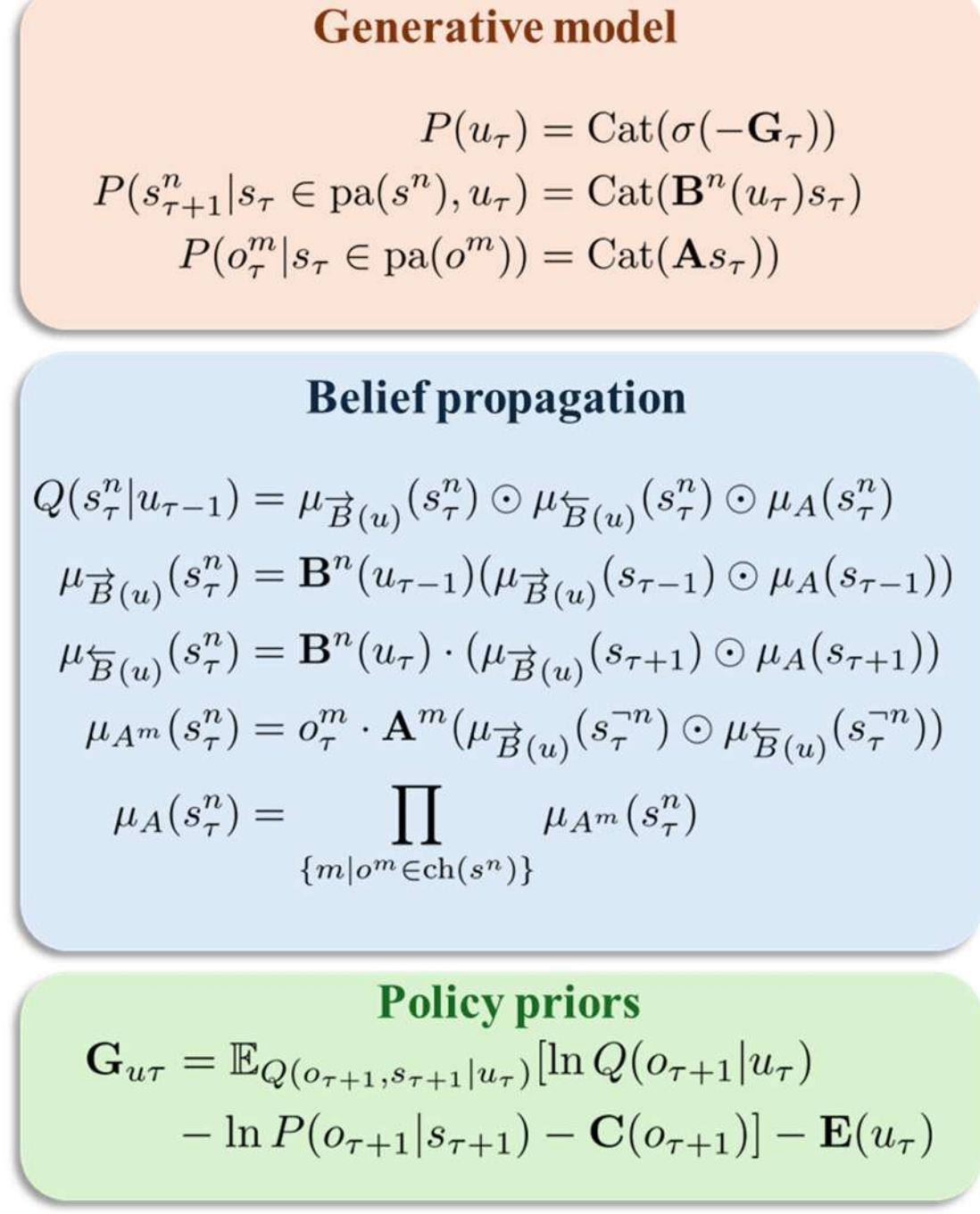

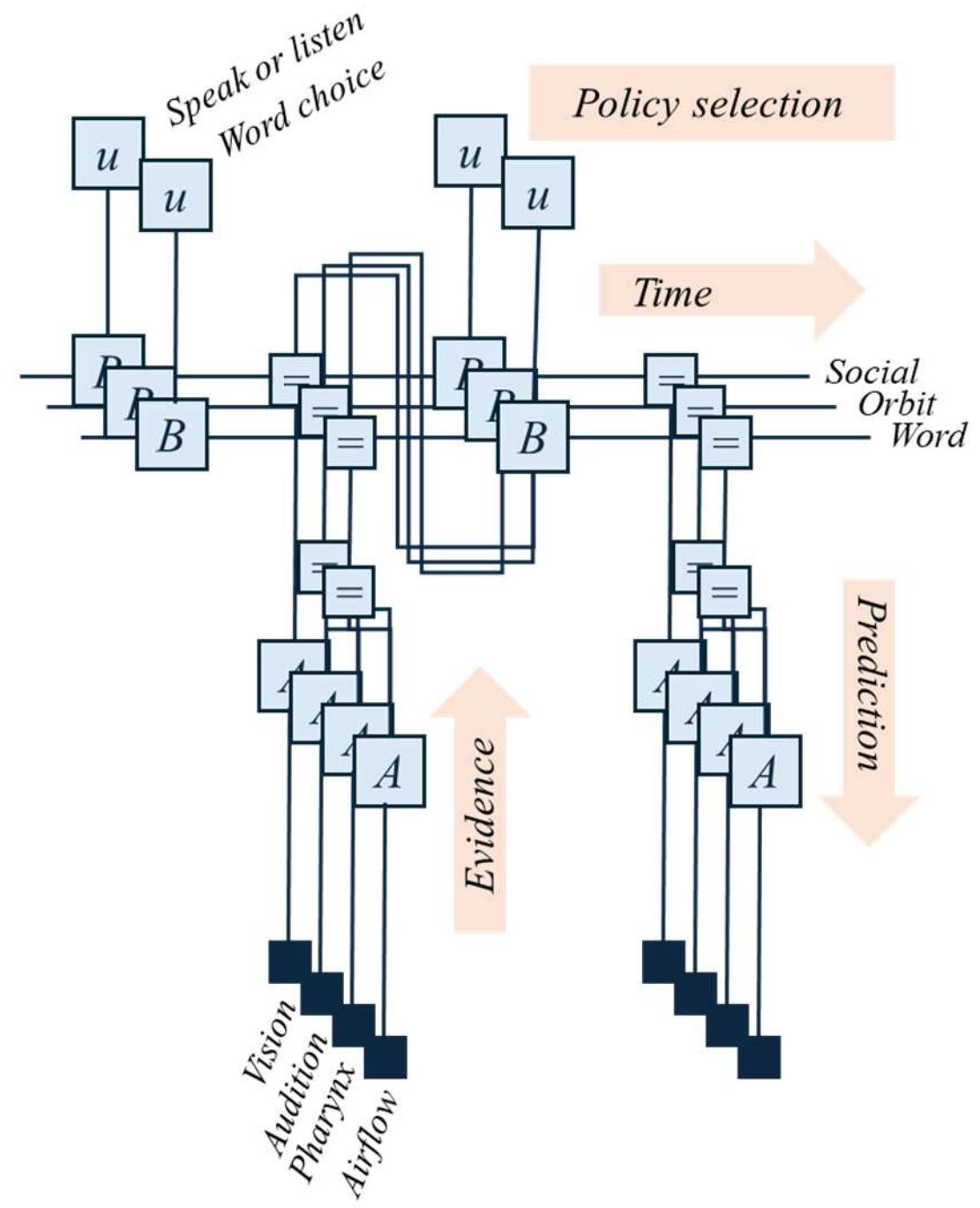

**Figure 1**

*Partially Observable Markov Decision Process*. This figure provides an overview of the form of the POMDP structure we use in this paper and the belief-propagation scheme used to solve it. The graph on the right uses a Forney factor graph notation (Dauwels, 2007; Forney, 2001; Loeliger et al., 2007) to detail the prior and conditional probabilities. Squares represent probability distributions, with edges (lines) representing variables. Boxes containing '=' are Dirac delta distributions that enforce equality between all connected edges. Filled boxes represent sensory data. The relevant distributions are expressed in the upper-left panel. The variable $u$ represents a controllable path (i.e., alternative transition probabilities under different actions), $s$ are hidden or latent states, and $o$ are sensory observations. Superscripts for these variables represent different factors or modalities. The notation *Cat* indicates a categorical probability distribution whose parameters are a matrix or vector. The σ indicates a softmax function. The notations 'pa' and 'ch' indicate the parents and children of a variable, respectively. For a probability distribution $P(a|\ b)$, $a$ is the child of $b$. The middle-left panel outlines the belief-propagation scheme used here. This is equivalent to a standard sum-product message passing scheme. Marginal beliefs $Q$ about any given edge are derived from the elementwise product of messages $\mu$ passed from each factor with which that edge connects. Messages from a given factor are derived recursively from other messages that converge upon that factor. The lower-left panel provides an expression for the prior over alternative paths. This is computed based upon the sum of a context insensitive prior weight **E** and an expected free energy comprising the average (expectation) of the difference in several log probability distributions. The **C** normally expresses value or preferences but is constant in all that follows. Rather than unpack this in detail here (see, e.g., (Da Costa et al., 2020)), we note that the only role the expected free energy plays in this paper is that it makes states associated with information gain more attractive and those states with greater ambiguity (e.g., those when both parties are talking) more aversive.

## 3—A generative model for speech fluency

In Figure 2, we illustrate the elements of our generative model. These are the hidden state factors and the (categorical) values each can take. The states comprise a set of (pseudo)words of variable number (we used a vocabulary of 16 words in the simulations that follow), a set of social contexts, and a position in a discrete orbit (which specified how many phonemes through the word we are) whose maximum length was 8 in keeping with the phonological length of the majority of English words (Marian et al., 2012). The words were generated by randomly combining a set of simplified phonemes common in English, with the constraint that the maximum length of a word was 7 phonemes, single phoneme words were vowel sounds, and each multi-phoneme word involved an alternation between vowel and consonant sounds (as defined below, noting that we neglect some of the subtleties in definitions of mixed and transition sounds for simplicity).

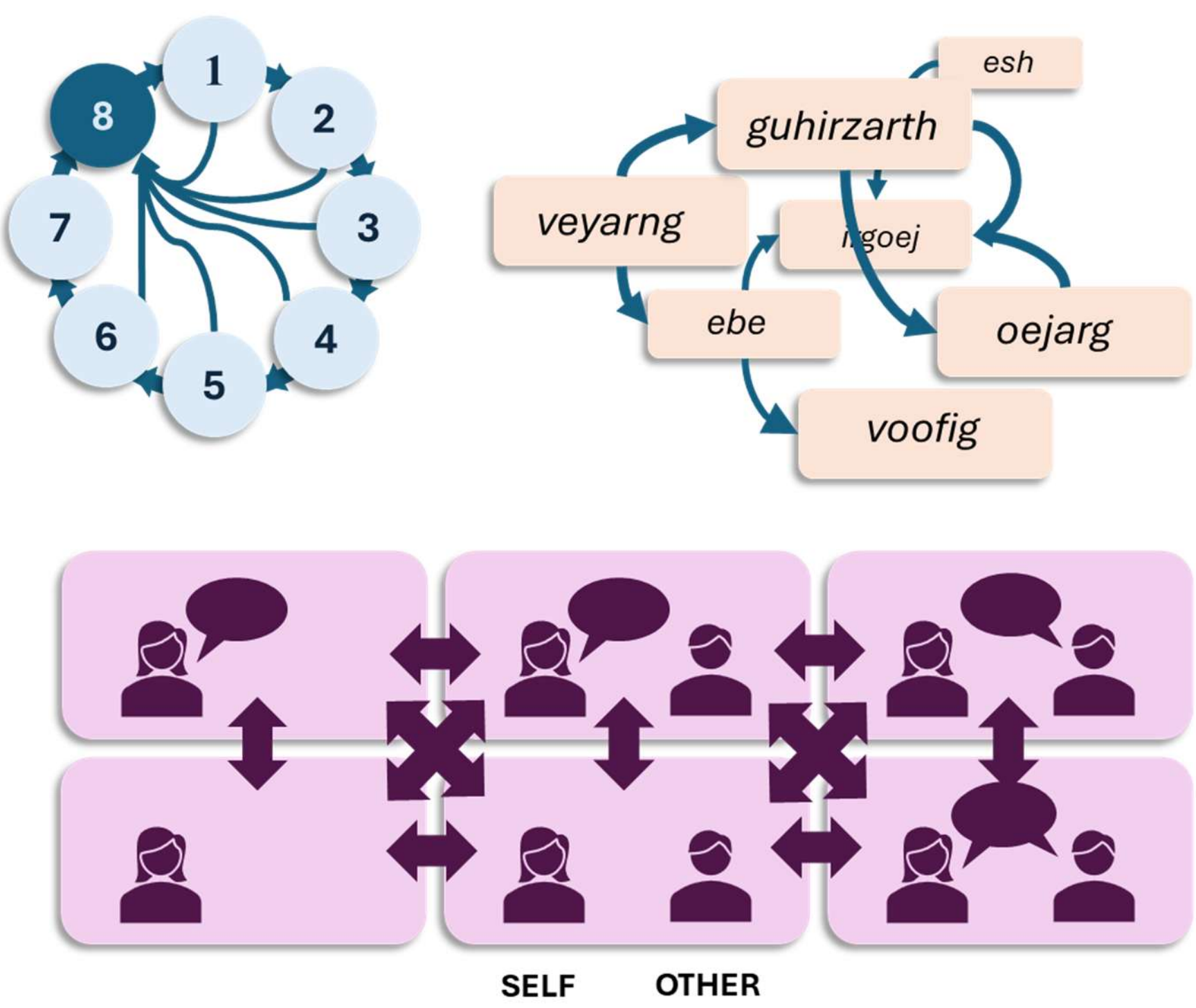

**Figure 2**

*States and transitions.* This graphic illustrates the sets of states and transitions that make up our generative model, and from which outcomes are generated. These states relate to the three horizontal lines in the factor graph of Figure 1. The upper-left states, shown in blue, represent the discrete orbit that determines how far we have progressed through a given word. The arrows show the allowable transitions between states, such that we progress round the orbit. Depending upon the current word, and the number of phonemes it has, the orbit may terminate early. This is indicated by the arrows representing transitions that skip to the $8^{th}$ state. For any given word, the orbit state predicts which phoneme should currently be audible, with the $8^{th}$ state consistently predicting silence. The upper-right states, shown in orange, deal with the words. The (pseudo)vocabulary is generated randomly as described in the main text. The transitions between words are constrained to be the identity transition unless the orbit state is 8, so that we can only progress to the next word if we have reached the end of the previous word. Otherwise, transitions are based upon several possible trajectories, each of which is constructed by randomly permuting the rows and columns of a matrix with ones along the leading diagonal. When the social context involves our subject speaking, the trajectory depends upon the current controllable path state. When it involves listening, the transition matrix is the average of those for all possible trajectories. The lower states, shown in purple, represent a social context, in which one can control transitions between states where they are or are not speaking, but not whether another person may arrive, leave, start speaking, or stop speaking.

A simple pseudo-syntax was generated by randomly permuting the rows and columns of a square matrix, whose dimensions matched the size of the vocabulary, with ones along the leading off diagonal. For the purposes of this paper, we generated 6 different such matrices. Treating each matrix as a transition probability matrix resulted in several possible repeating sequences of words. When the social context involved our agent speaking, they believed they were following one of these sequences of words. When they were listening, they believed their conversational partner made use of a transition matrix that was the average of the matrices for the alternative sequences, making the sequences less predictable.

One of the key differences here to most (perhaps all) previous POMDP Active Inference models is that we employ a much denser set of conditional dependences in the model of dynamics (c.f., the factorisation in (Mirza et al., 2016)). Specifically, we allow for the different factors (horizontal lines in Figure 1) of the hidden states to mutually influence one another. This means a transition probability matrix becomes a transition probability tensor. The reason for this becomes clear when we think about the relationship between word and orbit states. For the orbit, we know that the overall length of the orbit, and therefore the position at which we jump to the $8^{th}$ state, depends sensitively upon the word we are currently saying or listening to. This means the probability of going to the $i$-th orbit state from the $j$-th orbit state depends upon the $k$-th word currently in play. This probability is represented in the $ij$-th element of a matrix that is the $k$-th slice of a probability tensor.

The transition probabilities for all three sets of states were conditionally dependent upon other variables. The transitions for words depended upon both the position in the orbit and the social context. When the orbit state was anything but the $8^{th}$ state, the word transition matrix was an identity matrix, so that the word stayed the same until it was completed. In the $8^{th}$ state, if the social context meant being the speaker, the transitions were conditioned upon a controllable path variable. This allowed our agent to select between one of the syntaxes available, and to move to the next word in that syntax. If the social context involved someone else speaking, the transition probability was averaged across possible syntaxes. The transitions for the orbit state depended upon the current word being spoken. It involved a transition from 1 to 2, 2 to 3, and so on, until the length of the word (in phonemes) was reached, and then a transition to orbit state 8. Orbit state 8 always transitioned to state 1. This ensured the start of a new orbit coincided with the transitions from one word to another. Transitions in social context depended in part on an additional controllable path variable. This determined the transitions between whether our agent chose to speak or not. However, the transition probabilities additionally included non-controllable stochastic transitions in which the other speaker could arrive or leave and, if present, start or stop speaking.

This definition of the transition dynamics means all words start in the same orbit state but end on different orbit states depending upon their length. An alternative is to start words of different orbit lengths at different initial positions and work towards the same end state (i.e., for a word of length 3, start from orbit state 5 and then transition to 6, 7, and then the space on 8). Both formulations are available in the accompanying software demonstrations and result in qualitatively similar phenotypes. However, the sensitivities to different parametric lesions differ. While we will primarily report on the formulation outlined in Figure 2, where there is a common start position for all words, we will comment in the discussion on some of the differences seen when there is a common end position for all words.

The role of the 'orbit' state in our generative model offers an important point of contact to the role of perception-action cycles and intermittent sensory attenuation central to active inference accounts of motor control (Brown et al., 2013) and emphasised by Usler (2025) in their account of stuttering. Put simply, if movement is a form of prediction error resolution, then to initiate movement, one needs to suppress belief-updating (i.e., attenuate sensory precision) so that one can entertain the (transiently false) belief that one is moving, thereby inducing the prediction error that causes the movement. While we implicitly assume, in this discrete-time model, that sensory attenuation occurs between time-steps to allow for the action to change from moment to moment, this becomes particularly salient for the start of words. Usler's point that the right timing and rhythm of sensory attenuation is needed to allow for fluent speech coheres exactly with the idea that there are only specific points, during an orbit, in which one can change word—this is the break between words where there are no useful sensory data to determine what is being said, and all depends upon one's priors.

This set of hidden states determined the explanatory space available to the agent engaged in the simulation. The observations predicted under various combinations of these hidden states were (1) the auditory phoneme currently generated (depending upon the orbit state and the current word state); (2) two proprioceptive modalities, loosely related to articulatory and airflow sensations (also depending upon the orbit and current word state); and (3) a visual modality that reported the presence or absence of a conversational partner (which depended only upon the social context state).

The two proprioceptive modalities and phoneme were related as shown in Tables 1 and 2. These are simplified from[2] the concatenated columns and rows one might encounter in International Phonetic Alphabet (IPA) tables for vowels and consonants (International Phonetic Association, 2020). These are expressed using Latin letters—rather than more standard IPA symbols—for ease of compatibility with text generation via MATLAB figures. We will call the proprioceptive outcome comprising the concatenation of the row names in Table 2 and 3 the 'airflow' outcome, and those comprising the concatenation of rows the 'pharyngeal' outcome, acknowledging that both names are gross oversimplifications and that the sensory data resulting from such outcomes will combine proprioceptive and somatosensory information. This formulation means that every phoneme has a unique auditory outcome and a unique combination of two proprioceptive outcomes. As each combination of word and orbit states implies a specific phoneme, the tables directly define the '**A**' mappings from Figure 1 when the agent is speaking. For a given word and orbit state, these mappings predict three things: (1) the auditory phoneme, (2) the 'airflow' outcome, and (3) the 'pharyngeal' outcome corresponding to the column and row identities of the associated phoneme in these tables. Crucially, this also defines a set of combinations of proprioceptive outcomes—the empty cells—for which no meaningful auditory phoneme is generated. This is key in that it allows for articulatory errors that do not produce a phoneme compatible with English. In the auditory domain, mismatches between the phoneme heard and the proprioceptive outcomes associated with that phoneme play an important role in determining who is speaking. In other words, the proprioceptive signatures of different phonemes can overlap. Proprioceptive outcomes are generated only when our agent is the speaker. When both parties are speaking, the auditory outcomes (in our agent's generative model) are assumed to be generated from a uniform distribution.

[2] Specifically, and we note there are alternative classifications (for instance, some would define the post-alveolar phonemes as palatals), they are adapted from the tables presented in: Wikipedia contributors. (2026). *English phonology*. Wikipedia, The Free Encyclopedia. Retrieved 16/02/2026 from https://en.wikipedia.org/w/index.php?title=English_phonology&oldid=1335940802

There are, of course, alternative ways we could have formulated some of these elements. The rationale for using phonemes close to those seen in English is partly for illustrative purposes. In principle, we could have replaced these with labels $p1$, $p2$, $p3$, etc., and the following analysis would have been the same. However, each of these would still have to map to distinct proprioceptive outcomes. Using IPA-like phonemes gives us a simple and biologically valid way to do this. It also raises the possibility of generating audio waveforms, simply by passing the phonemic spellings through standard speech-synthesis software[3]. We have chosen not to constrain the space of words to real English words to avoid giving the misleading impression that this generative model is capable of meaningful word-level semantic or syntactic structures, reserving this for future work.

Furthermore, the factorisation of proprioceptive outcomes into 'airflow' and 'pharyngeal' modalities is designed to reflect the coordination of different types of motor effectors in generating speech. We could have combined these (technically, using their Kronecker products) into a single list of alternative outcomes. This would be mathematically equivalent from the perspective of inference. However, it does introduce an interesting dimension to the generation of speech. Specifically, sampling a single combined proprioceptive outcome versus sampling the two elements independently makes little difference when the proprioceptive outcomes are highly predictable. This is because, when we know which word we want to say, and when there is a clear mapping to the current phoneme, the relevant pair of proprioceptive outcomes are predicted, and therefore sampled, with high precision. However, when there is greater uncertainty, it is possible to sample a pair of proprioceptive states that are incompatible with speech (i.e., an empty element of either Table 2 or 3). This captures, in simplified form, the complex motor coordination required for speech. To generate meaningful phonemes requires careful coordination of, for instance, the appropriate airflow and articulation, and a mismatch between the two might lead to a failure to generate the intended sound.

**Table 2 - Consonants**

| | **Labial** | **Dental** | **Alveolar** | **Post alveolar** | **Palatal** | **Velar** | **Glottal** |
|---|---|---|---|---|---|---|---|
| **Nasal** | m | | n | | | ng | |
| **Plosive fortis** | p | | t | ch | | k | |
| **Plosive lenis** | b | | d | j | | g | |
| **Fricative fortis** | f | th | s | sh | | x | h |
| **Fricative lenis** | v | eth | z | zj | | | |
| **Approximant** | | | l | r | y | w | |

[3] For interested readers, this is implemented in the associated MATLAB demos (see software note). However, full exploration of this is reserved for future papers.

**Table 3 - Vowels**

| | Front short | Front long | Central short | Central long | Back unrounded | Back rounded short | Back rounded long |
|---|---|---|---|---|---|---|---|
| **Close** | I | i | oo | u | | | owe |
| **Mid** | e | e | ir | oe | | or | |
| **Open** | æ | | | | ar | | |

## 4—Simulating fluent speech and successful speech segmentation

In this section, we present examples of fluent speech and successful speech perception. The parameter settings used as defaults for these simulations are given in Table 4. These are expressed as precision parameters. Precision can be understood as a kind of confidence. This is not in the lay sense of 'self-confidence' but in the statistical sense that there is a narrow range of plausible values that a random variable might take. While useful to think of precision in terms of confidence, it is important to make this distinction in accounting for self-report findings in people with dysfluency, including stutter, who (almost by definition) (Neef & Chang, 2024) report they know exactly what they want to say. There is a complex relationship between precision and the subjective (or metacognitive) experience of confidence—with the two often dissociable from one another. There is an active research field, predicated upon signal detection theory, that seeks to understand these relationships (Barrett et al., 2013; Fleming & Daw, 2017; Fleming et al., 2010). In this paper, we remain agnostic about the subjective feeling of confidence and will focus only upon precision in the statistical sense.

The choices of these parameter values are largely arbitrary, reflecting a high degree of precision at baseline, while allowing for the possibility of changes in speaker (with interruption being less likely than speech initiation when silent). We would expect variations in the default parameters to have an impact on, for instance, the thresholds at which perturbations result in behavioural changes. This is something we will explore towards the end of the paper, in which we compare the interactions between pairs of parameters. The lower-case symbols in the table correspond to elements of the upper-case symbols in the distributions in Figure 1. Note that all parameters are treated as probabilities in the 0-1 range except for the $e_1$ parameter, which plays the role of an inverse temperature. This is because we treat the distribution of possible word trajectories that are likely to be selected as the softmax of a randomly generated vector (from a normal distribution with spherical covariance), with one number per trajectory. In all cases, we start with uniform prior beliefs about the word currently being spoken. When our agent should be speaking, we set a prior belief that the controllable path for the social state towards speaking ($e_2$) with probability 1. When listening, we set this probability to be 0. In some of the simulations, we initialise the environment to include a second person, and in some simulations, we initialise it such that the second person is speaking from the start.

**Table 4 – Default parameters**

| Parameter | Value | Interpretation | Range |
|---|---|---|---|
| $a_1$ | 1 | Precision in prediction of phonemes given latent states | [0 1] |
| $a_2$ | 1 | Precision in prediction of proprioceptive outcomes given latent states | [0 1] |
| $b_1$ | 0.1 | Probability someone will arrive if I am alone | [0 1] |
| $b_2$ | 0.1 | Probability someone will leave if I have company | [0 1] |
| $b_3$ | 0.8 | Probability someone will start speaking if I am silent | [0 1] |
| $b_4$ | 0.4 | Probability someone will stop speaking if I am silent | [0 1] |
| $b_5$ | 0.4 | Probability someone will start speaking if I am speaking | [0 1] |
| $b_6$ | 0.8 | Probability someone will stop speaking if I am speaking | [0 1] |
| $b_7$ | 1 | Probability of most likely next orbit state given current state | [0 1] |
| $e_1$ | 8 | Precision (inverse temperature) in what I want to say | $> 0$ |
| $e_2$ | 1 or 0 | Probability that I will start speaking | [0 1] |

The upper third of Figure 3 shows the output of running a simulation using the generative model of Figures 1 and 2 in an environment where our agent is on their own, and where they hold precise prior beliefs that they will speak and know what they want to say (i.e., which syntactic path they are on). The plots show a pictorial representation of the outcomes and beliefs available at the final time-step of the simulation (on the left) alongside the set of probabilistic beliefs formed at each time-step with the outcomes presented (on the right). The social state is quickly inferred to be the 'speaking to myself' state. In the absence of auditory or proprioceptive outcomes in the first two steps, an inference is made that it is likely the $8^{th}$ orbit state. This prompts a belief at the next step that it should be the start of a word, with several words inferred to be plausible. This is enough to predict and generate a consistent phoneme, with a second generated at the next time-step. From here on, the simulated speaker is confident about their word choices and generates consistent auditory and proprioceptive data.

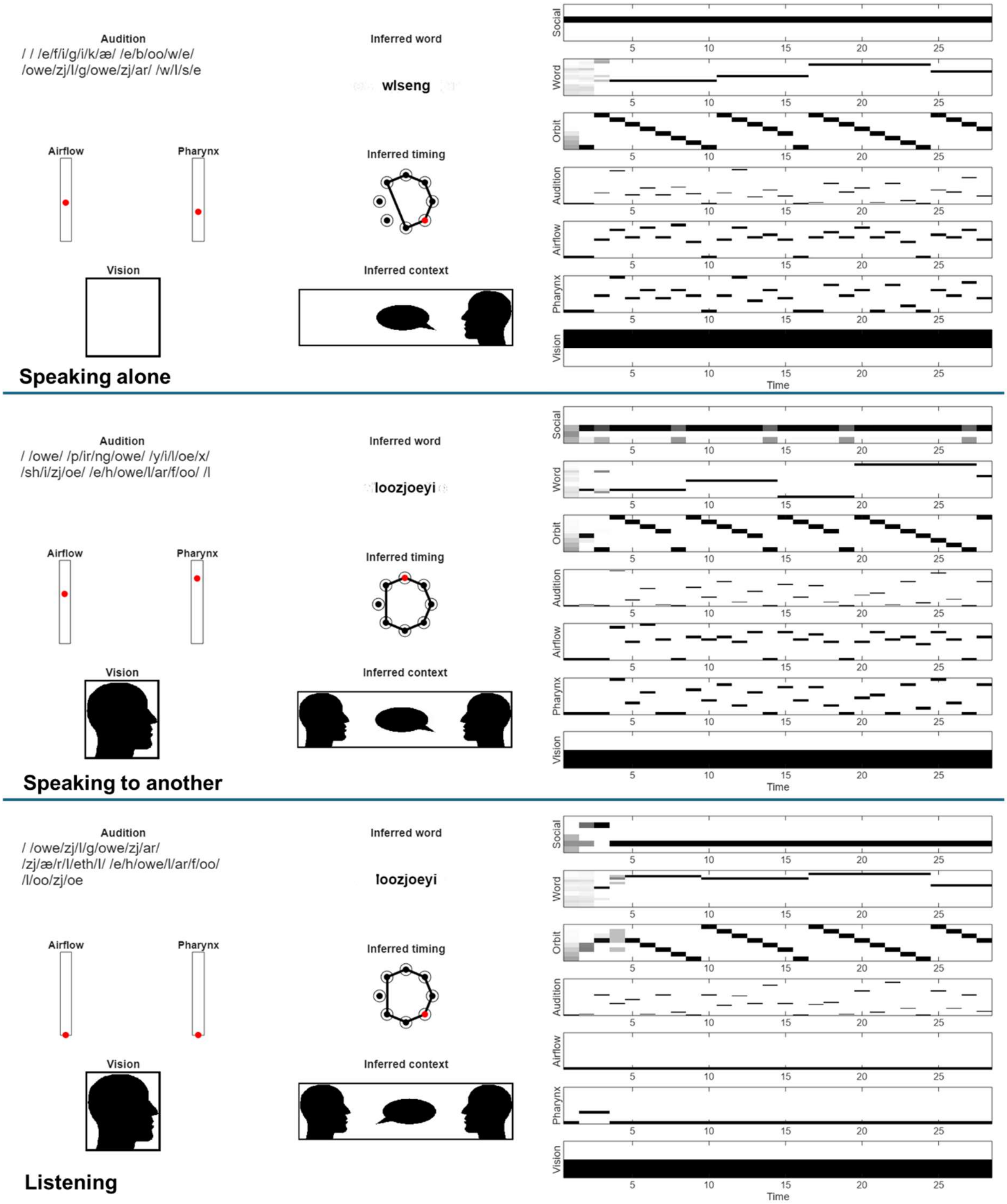

**Figure 3**

*Behaviour and perception under default parameters.* This figure reports three simulations in differing contexts under the parameter settings detailed in Table 4. For each simulation, we present a graphical representation of the sensory data and inferences on the left, and a plot of beliefs and observations over time on the right, which can be interpreted much like the 'raster plots' one might encounter in single-cell electrophysiology research. Each row of each plot represents an alternative categorical value. The graphics on the left, for each simulation, show the final frame of the animation that is generated by running the associated MATLAB demo (see software availability note). This means that it represents the state of the simulation at the final time-step. The 'audition' graphic shows the sequence of phonemes generated so far (these are presented one at a time). This corresponds to the 'audition' raster plots on the right, which show this sequence of auditory phonemes over time. In the raster plots, the 'silent' state is the lowest row. The 'airflow' and 'pharynx' graphics indicate different proprioceptive data with red dots at different heights. The heights of the red dots correspond to the positions in the corresponding rows in the raster plots on the right, with the lowest row being a resting state. The 'vision' outcome is presented as either an empty frame (nobody there) or a silhouetted figure (another potential speaker present). The corresponding raster plot's first and second rows represent these outcomes, respectively. The 'inferred word' graphic displays all possible words, with font colour weighted by the beliefs shown in the 'word' raster plot. The 'inferred timing' graphic shows beliefs about the current orbit state in red, with the inferred orbit (i.e., word length) shown with black links. This is derived from the 'word' and 'orbit' beliefs shown on the right. Finally, the 'inferred context' graphic expresses the 'social' beliefs, with the silhouette on the left shaded according to the belief that someone else is present, and the speech bubble shaded according to beliefs about who is speaking. A description of the key findings from these simulations is detailed in the main text.

The middle third of Figure 3 shows what happens when we place another (silent) person in the environment. Here, we see some subtle but important changes to the agent's beliefs. First, although they mostly believe they are in a social context that involves speaking to another, during the pauses between words there is a brief window in which they believe the other might start speaking—leading to the 'both speaking' outcome. This seems sensible, as the pauses could be consistent with either person speaking. However, the clear relationship between audition and the proprioceptive outcomes associated with a phoneme while a word is voiced are sufficient to ensure our agent can infer with precision that they are the one speaking.

The lower third of Figure 3 shows what happens if we allow the other speaker to talk and give our agent prior beliefs that they are unsure of the syntactic trajectory they would adopt if they were speaking and that they do not expect to start speaking. Here, we again see some confusion at the start, but by time-step 5 there is a precise inference that our agent is listening to another, and an effective segmentation of the words and orbit lengths. This demonstrates the plausibility of using the same set-up and model for both speech production and perception.

## 5—Possible mechanisms for breakdown in fluency

There are many possible mechanisms we could consider for speech fluency breakdown, and the discussion here is only really scratching the surface. However, we will set out some of the broad mechanisms that may be relevant and show how one could develop empirical predictions from these. The broad categories of mechanism are as follows (with the affected probability distributions in parentheses):

- Loss of precision in the word sequence ($e_1$)
- Loss of precision in whether to speak ($e_2$)
- Loss of precision in when to speak ($b_7$)
- Inappropriate social priors—e.g., expectation of being interrupted ($b_5$)
- Inappropriate weighting of proprioceptive and auditory sensations ($a_1$)

Our focus in what follows will be the ways in which we would expect observed behaviour to differ according to these mechanisms. We start from the first two manipulations, which both relate to controllable paths or policies. As these priors relate to things that our simulated subject controls themselves, we can see the effects of these in the absence of another person (we will return later to the modulation of these effects by social context).

***Loss of prior precision over word sequence***

The simulation reported in the upper half of Figure 4 attenuates the precision in the upcoming sequence of words, while the lower half attenuates the precision in whether to start or stop speaking. As is evident from Figure 4, the patterns of speech generated under these lesions are qualitatively different. When the precision is low for the word sequence, we see two features characteristic of breakdown in speech fluency. These are the increased gaps between words (seen by multiple empty phonemes '/ /' between words in the auditory outcome) and repetition of phonemes at the initiation of a word ('/e/f/ / /e/f/…'). We also see repetition of the entire word given the lack of certainty about which word to choose next. Both full and partial word repetitions have been described as features of stuttering (Gregg & Yairi, 2012; Howell et al., 2006), as have delayed initiation of words (Tetnowski et al., 2021). Long gaps between words, as part of a 'halting' speech pattern, have also been described in non-fluent aphasias (Grossman, 2012)—although this is not to suggest an equivalence between stuttering and non-fluent aphasias.

That loss of prior precision over the word sequence appears to reproduce some features of stuttering offers a possible explanation for an interesting empirical finding. This is the demonstration that short delays in auditory feedback, of the order of up to about 100ms, can be used to induce fluent speech in people who stutter (Goldiamond, 1965; Kalinowski et al., 1996; Lincoln et al., 2006), and neurophysiological correlates of this (Daliri & Max, 2018). The relevance of this when there is greater uncertainty about *trajectories* is that delayed auditory feedback—combined with current proprioceptive feedback—ensures one always has access to two points along the trajectory, repairing some of the uncertainty induced by low prior precision. Intuitively, seeing two adjacent frames of a ball flying through the air gives one much more information about its path than a single frame. Similarly, information about two adjacent phonemes tells us much more about a verbal trajectory, even if these come from two different sensory modalities.

***Loss of precision in whether to speak***

The lower half of Figure 4 shows a different pattern. Here, the loss of precision in whether to start or stop speaking leads to gaps between words but, rather than starting each word later, the subject simply skips the first phoneme of each word. This is evident from the plots on the right, in which the first state of the orbit is often accompanied by a silent auditory outcome (lowest row). Interestingly, the silent auditory outcome is sometimes accompanied by the absence of movement in the pharynx or airflow outcomes (again, the lowest row for each). Specifically, some pharyngeal movement (e.g., time-step 11) occurs in the absence of the necessary airflow to generate a sound, suggesting a degree of discoordination in the process of articulating speech when unsure whether to speak. One could interpret this as an insufficiently precise proprioceptive prediction or command to the relevant muscle groups to jointly activate the right set of actions when one is unsure whether they expect to be speaking. This combination of discoordination and phonemic omissions or deletions have been observed in speech apraxia syndromes (Code, 2021; Ogar et al., 2005; Peach & Tonkovich, 2004).

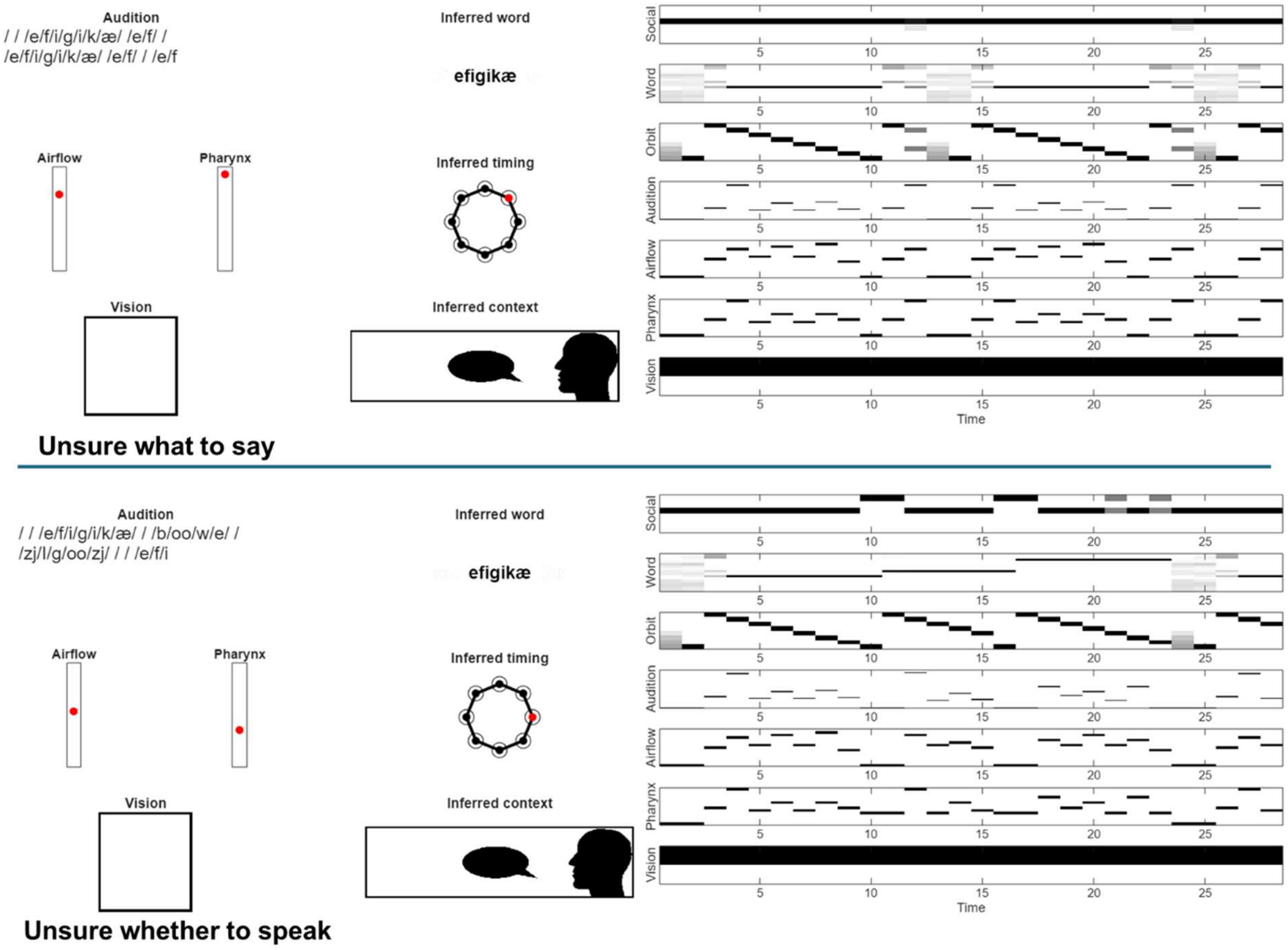

**Figure 4**

*Uncertainty over controllable paths.* This figure makes use of the same format as Figure 3. Please refer to the corresponding caption for details. In the upper panel, the inverse temperature was set to ¼ for the controllable path through word space, effectively flattening the distribution of possible paths. This results in a high degree of uncertainty at the start of each new word, resulting in prolonged intervals between words with phoneme and entire word repetitions. In the lower panel, the precision of whether to speak was set to ½. This led to a different pattern of speech disturbance, with prolonged intervals between words secondary to omission of the first phoneme of some words. Additionally, during these omissions, we see discoordination of the articulators as indexed by absent airflow in the presence of pharyngeal movement.

### *Loss of precision in when to speak*

Next, we will consider the effects of uncertainty in the orbit transitions. Here, because the orbit is shared between our subject's inferences about their own speech and of the other person, we need to consider the effects of altering the social context and so present simulations both when speaking and listening. When speaking to oneself, speech is well preserved when the precision in transitions ($b_7$) is a probability of 0.8 or above. Speech is completely extinguished if we drop this precision to 0.7. Below this value, the subject is silent throughout and unable to *initiate* speech. This is also true in the presence of another person as shown in the upper half of Figure 5.

Interestingly, our simulated subject has limited difficulty in segmenting the speech of another person at this level as shown in the lower half of Figure 5, once sufficient evidence has been accumulated about the orbit state in play. Intuitively, this is because hearing a phoneme /ar/, when one's vocabulary contains a word with the phonemic sequence /owe/zj/I/g/owe/zj/ar/, provides evidence for the 7$^{th}$ orbit state that helps to counteract the accumulated uncertainty in orbit states resulting from loss of precision in transitions. The phenotype here shares features with expressive (Broca's) non-fluent aphasia syndromes (Grossman, 2012) in that there is relative preservation of speech comprehension despite loss of the ability to initiate meaningful speech. Interestingly, there is evidence that those with such syndromes can sometimes sing with relative fluency (Hébert et al., 2003). Explanations for this finding range from slower rates of phoneme articulation in singing compared to speech, changes in hemispheric dominance, to a restriction of the available vocabulary based upon learned songs. Another plausible explanation here is that the rhythmic information associated with music provides strong priors about when transitions occur in time, about the beat identity within a bar (which might be tied to orbit position), or that melodic structure provides cues that help to resolve the orbit position which might repair some of the loss of precision about orbit states under the computational lesion presented here. A melody might reduce uncertainty at branch points and removes uncertainty about who is speaking. This may explain the promise shown by the use of Melodic Intonation Therapy, which uses musical approaches including singing and rhythmic tapping in rehabilitation for nonfluent aphasia (Albert et al., 1973; Norton et al., 2009; Popescu et al., 2022).

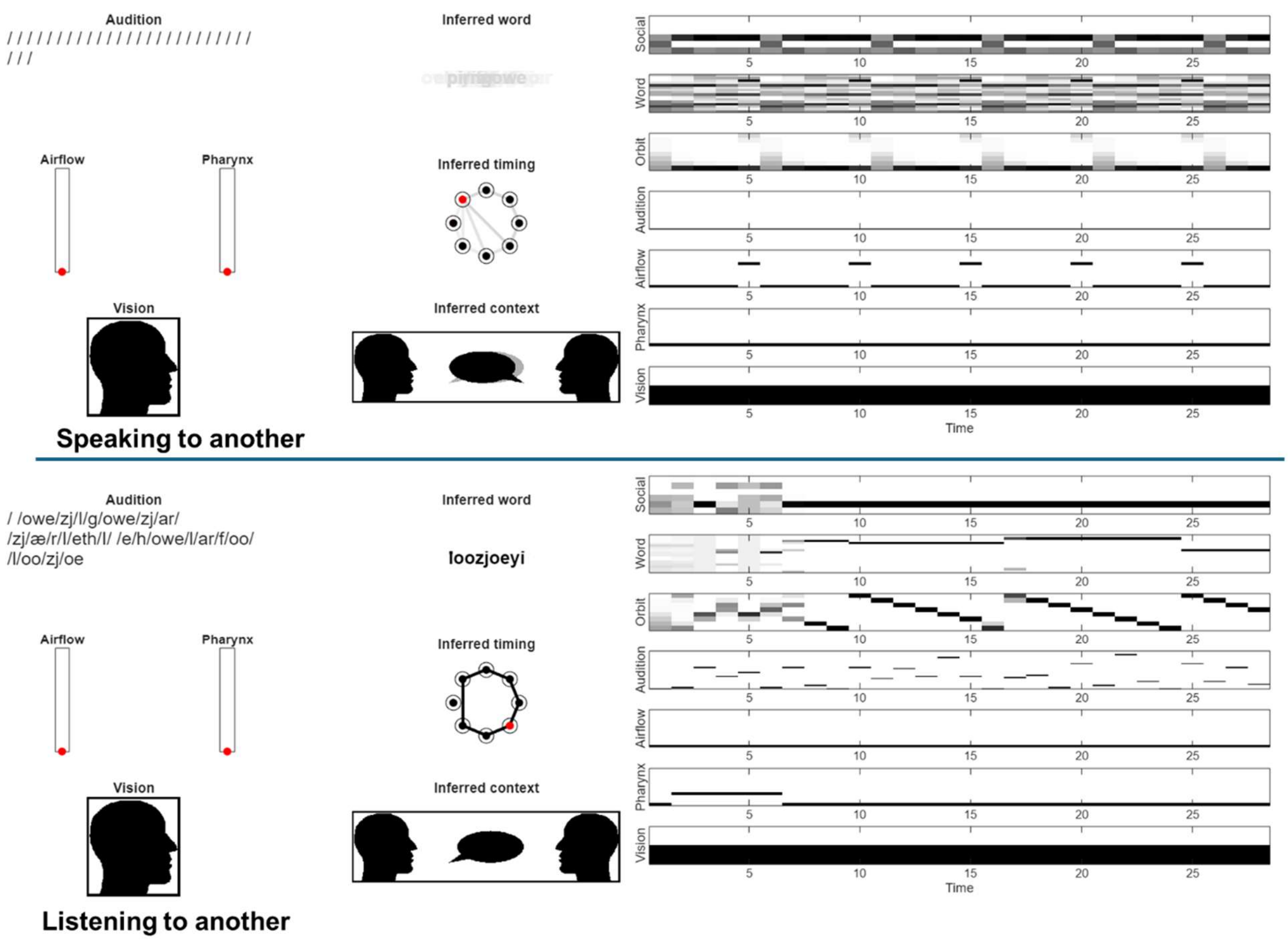

**Figure 5**

*Uncertainty in transitions.* This figure, using the same format as Figure 3, shows the result of reducing precision in transitions around the orbit. This means the subject has a belief that they may end up transitioning to the orbit state they expected, but there is a small probability of ending up in one of the adjacent orbit states. The result is impaired initiation of speech, but relative preservation of speech perception—although note it takes longer to confidently infer the word than in previous simulations.

### *Inappropriate social priors—e.g., expectation of being interrupted*

Changes in prior beliefs about the social transition probabilities and specifically changes in beliefs about the probability of being interrupted, might be expected to cause interference between the speaker and their anticipation of another speaking. We can simulate this by increasing the perceived probability that someone will start speaking when our subject is speaking. Interestingly, with all the other parameters set as in Table 4, this did not cause any impairment in speech fluency or word recognition. As this is not appreciably different from the behaviour shown in Figure 3, we do not include an additional figure to display this here. The implication is that a belief that one is likely to be interrupted, on its own, is not sufficient for breakdown in fluency.

### *Inappropriate weighting of proprioceptive and auditory sensations*

Figure 6 shows what happens when we manipulate the precision of the auditory predictions (i.e., make the mapping from word and orbit state to the expected sound more ambiguous). By decreasing the expected probability of the most likely phoneme from 1 to 1/16, and distributing the remaining probability amongst the other 36 phonemes, we see a few interesting changes. First, when speaking, there is a higher probability assigned to the possibility that both parties might be speaking (seen as grey shading of the lowest row in the beliefs about social states). However, this does not interfere with speech generation. This is because the proprioceptive states remain informative and predictable. More strikingly, when listening to another person, there is a loss of precision in the spoken word. While, at most time-steps, the true word is inferred to be the most probable, there is much more uncertainty associated with this. Qualitatively, this behaviour resembles receptive (Wernicke's) aphasia in which an interpretation of an auditory stream as meaningful words is lost, despite relative preservation of speech generation (Buckingham & Kertesz, 1974; Damasio, 1992).

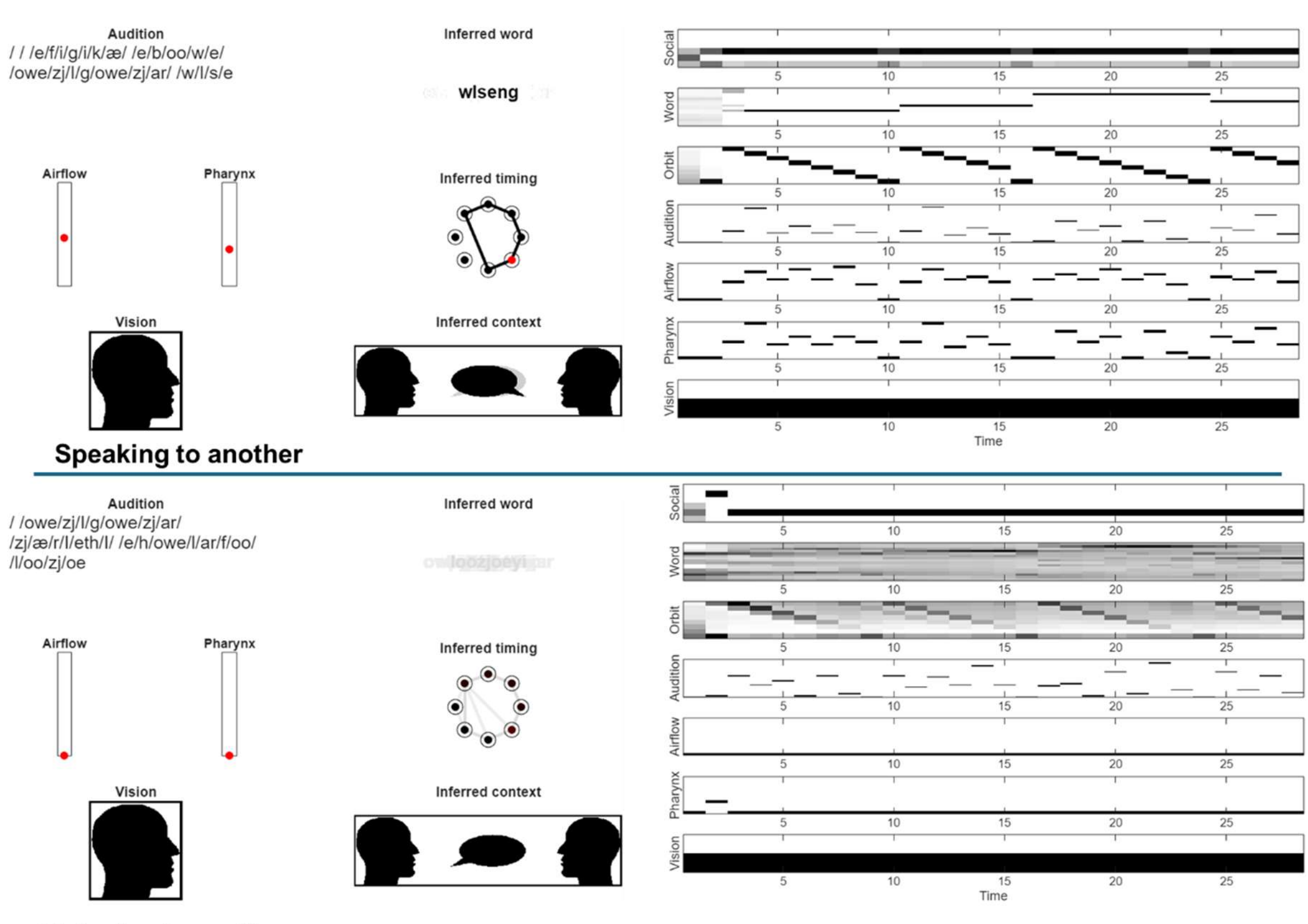

**Figure 6**

*Uncertainty in phoneme prediction.* This figure, using the same format as Figure 3, illustrates the effect of reducing the precision with which auditory phonemes are predicted from the orbit and word states. When speaking to another, this causes limited impairment in speech generation as the proprioceptive outcomes are sufficient to provide feedback on what is being said. However, when listening, our simulated subject becomes much less confident both about the word currently in play and the orbit states.

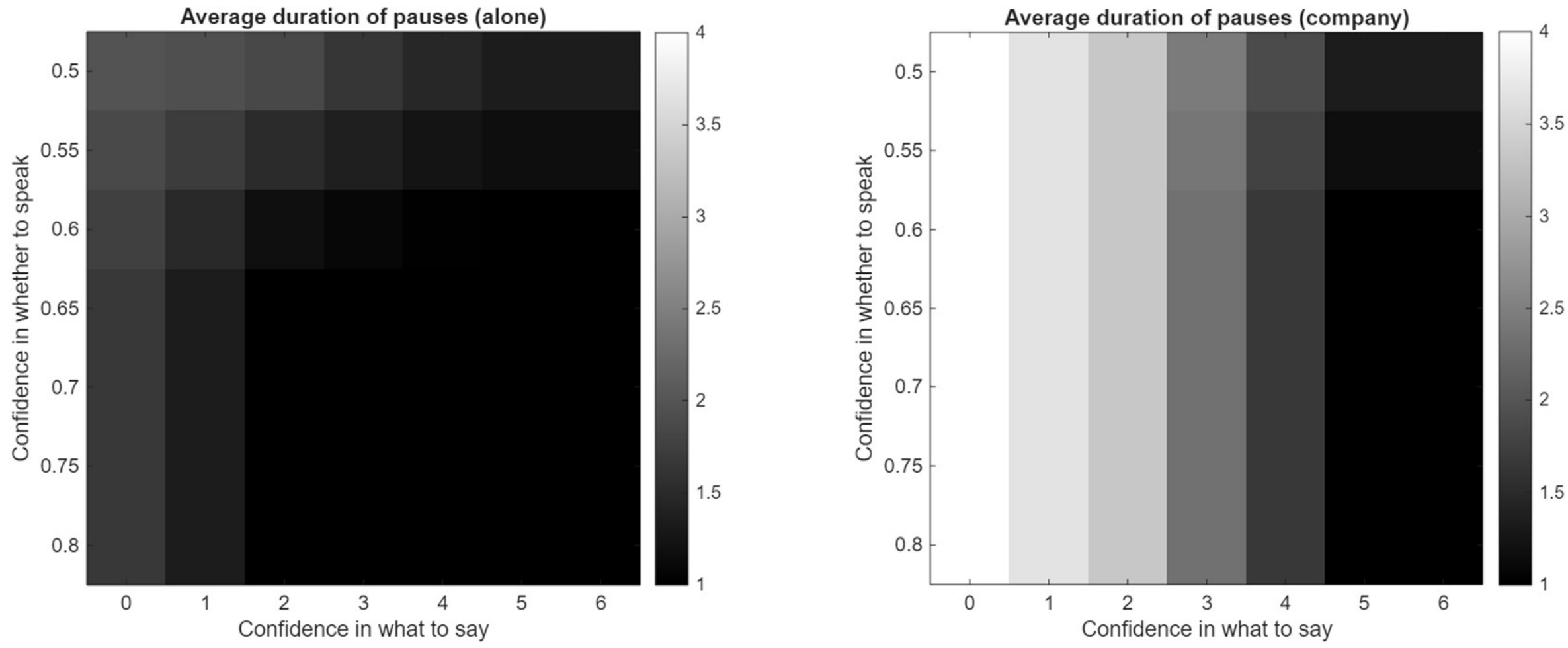

**Figure 7**

*Interactions of parameters and social context.* This figure illustrates the effects of manipulating the $e_1$ ($x$-axis) and $e_2$ ($y$-axis) parameters in the context of speaking to oneself (left) or speaking to another silent party (right). The shading of each element gives the average length of pause between words in a 32 time-step period. These simulated data have been smoothed using a 3x3 moving average kernel. Note the symmetry of the left plot across the diagonal, with the symmetry broken in the right. The implication is that, if a fluency deficit is a result of changes in these two parameters, one should be able to recover these parameter values from any given measurement of the average lengths of pauses between words in the presence or absence of conversational partners.

A final simulation result is shown in Figure 7. This is to illustrate the importance of interactions between parameters and social settings. In constructing these figures, we simulated 32 time-steps in the presence or absence of a second person who, when present, remained silent throughout. We repeated this for different values of the two $e$ parameters from Table 4. For each combination of parameters and social context, we computed the average length of the gap between words, excluding any pause before the first word is spoken. This is a coarse measure, which includes phoneme omissions as well as 'halting' and speech 'blocks.' However, it makes clear empirical predictions about the speech characteristics one would expect given the computational lesions that might underwrite a dysfluency syndrome. First, we see that there is an approximately symmetric pattern of deficit when speaking to oneself with severity of parameter manipulation (left panel). However, this symmetry is broken once we introduce another person, with relative invariance in the effect of precision in whether to speak. Such plots also speak to the possibility of compensatory strategies. Mild deficits in precision in whether to speak, or in the word sequence, could be compensated for by increasing precision in what to say, or whether to speak, respectively. For instance, a well-rehearsed script might help someone, with a mild deficit in precision about whether to speak, to engage in fluent speech.

## 6—Computational Neuroanatomy

One of the ways in which models of this sort can be used is in developing regressors for functional imaging that can help to test hypotheses about the computational anatomy. By simulating the belief-updating that corresponds to a given behavioural output, and associating beliefs about subsets of variables with specific sets of voxels, one can assess whether the trajectory of the belief-updates (suitably convolved with response functions) is predictive of activity in those voxels (see, e.g., (Nour et al., 2018; Schwartenbeck et al., 2015)).

It is important to think about the anatomical hypotheses one might wish to test, and the constraints the model itself places on the network architecture (Parr & Friston, 2018). Typically, the strongest constraints are the sensory and motor elements. We expect auditory (phonemic) input to involve the superior temporal gyrus (Liebenthal et al., 2005; Price et al., 1996), and motor output to act via motor cortical projections to brainstem nuclei and cervical spinal motor neurons (Kearney & and Guenther, 2019; MacLarnon, 2011). From here, we have a set of constraints based upon the conditional dependencies in the probabilistic generative model. Figure 8 shows one plausible arrangement of this network of connections. This arrangement is partially motivated by the pattern of deficits we saw in the simulations above under alternative lesions and partly by previous functional imaging and anatomical findings (Hickok, 2009b). For instance, there is evidence for a left posterior Sylvian region that is responsive to speech perception and production (Buchsbaum et al., 2001) that is a good candidate locus for beliefs about words regardless of who has spoken them. The subcortical structures involved in selecting between the alternative policies have been omitted for visual clarity but would be expected to include cortico-basal-ganglia loops, which notably have a role in acquired (Theys et al., 2013; Theys et al., 2024) and developmental (Alm, 2004; Chang & Guenther, 2020) stuttering. It is here that we might expect changes in the *e* parameters from Table 4 that deal with alternative courses of action.

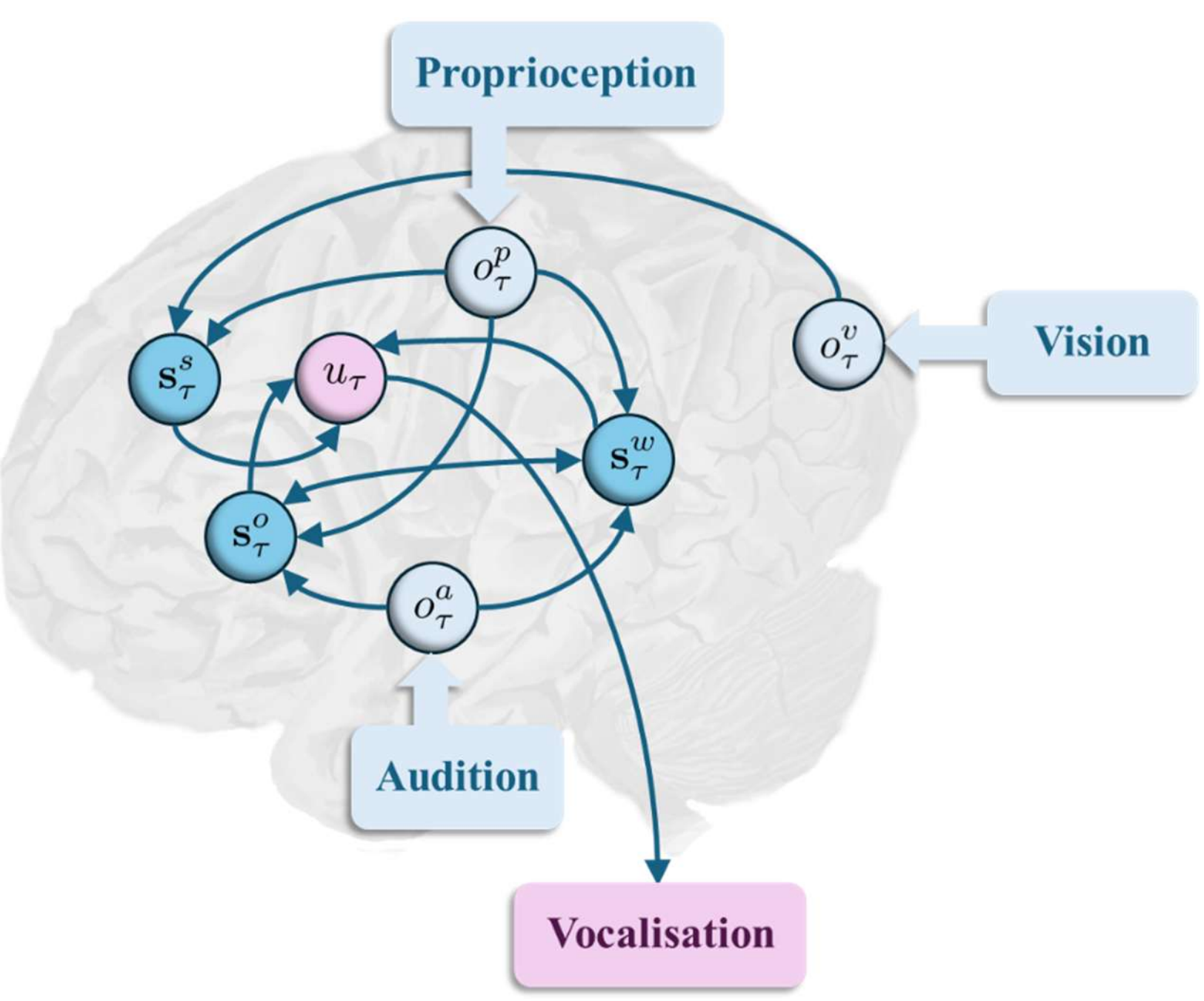

**Figure 8**

*Computational neuroanatomy of speech fluency.* This figure presents a plausible computational anatomy for speech fluency as mediated by the Active Inference model we have presented. This is somewhat speculative and represents one of several hypotheses that could be advanced within the constraints of the computational architecture and pattern of lesion behaviours. Beliefs about states are represented as bold **s** symbols, with superscripts *o, s, w* standing for orbit, social, and word states, respectively. Observable data are expressed using the *o* symbol, with superscripts *a, p, v* standing for auditory, proprioceptive, and visual modalities, respectively. Proprioception here is the combination of pharyngeal and airflow modalities and likely also involves an element of somatosensation. The *u* variable represents the actions taken. The subcortical networks likely involved in evaluating alternative actions have been omitted here. All nodes are intended as relatively course representations of populations of neurons.

Lesion mapping in speech and language syndromes (Fridriksson et al., 2018) offers another set of constraints. As we saw in the simulations, receptive dysphasic patterns, of the sort associated with damage to superior temporal and posterior superior temporal regions, emerged when we lesioned the connectivity between auditory phoneme representations and word representations. Similarly, expressive dysphasic patterns, of the sort associated with lesions to the inferior frontal gyrus and its connections emerged when we introduced uncertainty into transitions between orbit states. The functional interpretation of the relationship between the frontotemporal parts of the network in Figure 8 cohere with ideas about the structuring of words into sentences (Friederici, 2002), in that our 'frontal' orbit states are essential in segmenting words and timing the progression onto the next word. They also fit with evidence that, during more explicitly rhythmic activities, like singing, premotor (precentral) regions are specifically engaged (Zarate & Zatorre, 2008).

Our model encapsulates sensory speech perception and motor speech generation within a single framework. It is worth noting that one of the obstacles in the motor theory of speech perception is how to account for the double dissociation between receptive and expressive aphasias (Meister et al., 2007). Surely, if speaking and listening share mechanisms, then they should not be dissociable. Remarkably, active inference shows how these functions can be differentially impacted by specific lesions within the model.

## 7—Potential extensions and predictions

We have deliberately formulated the model in a way that can be straightforwardly mapped to experimental manipulations. For example, a participant might speak while another person is present or absent as indicated by an image on a stimulus display screen. The probability that the other person arrives, leaves, and starts talking could be manipulated. Participants could be pre-trained on some simple scripts and cued with different levels of certainty about the script they should be speaking from, if at all, at different points in time. Previous functional imaging experiments have implicitly manipulated precision about what to say in people who stutter using tasks such as sentence reading (highly constrained) versus describing a picture (less constrained) (Demirel et al., 2024). Our theoretical analysis suggests there may be value in looking at the interaction between these conditions and the presence or absence of another potential conversational partner.

There are several predictions one could test using such a paradigm, including fluctuations in the ability to decode words at specific intervals with non-invasive physiological imaging, the interaction between progressive perturbation of parameters—such as we might expect in some neurodegenerative disorders—and the patterns of deficit both when speaking to oneself and speaking to another, and (based on Figure 8) the loci of activity we might expect to measure using parametric regressors in functional imaging.

As an example of the first of these, looking at the belief plots on the right of Figure 4, we see that during silence between words, and even during the first phoneme, there is a high degree of uncertainty about the word selected—something that might be amenable to assessment with a magneto/electroencephalographic decoding paradigms (Dash et al., 2020; Défossez et al., 2023). If a relative loss of precision in the word sequence underwrites breakdown of speech fluency, we would expect decoding accuracy of the current word in the sequence to drop in the pause between words and during the first phoneme of new words for participants with disruptions of speech fluency, relative to a neurotypical participant (e.g., upper plot of Figure 3), but be retained during the spoken words.

The model predicts many quantitative testable phenomena, which require further simulation to pin down in specific contexts (e.g., stuttering versus aphasic syndromes). For example, what are the critical features that lead to dysfluency? Inherent factors may include sentence branch points, variable-length words (Soderberg, 1971), predictability of phonemes, and the unpredictability of another person. Further, modifiable factors that limit stuttering, such as rhythm or auditory cues, number of other people, and the benefits of filled pauses or discourse markers, could be investigated and may generate clinically useful hypotheses.

A core issue that we have not addressed here, but which would be interesting to examine further, is the influence of the vocabulary itself. One might expect greater difficulties with vocabularies in which many different words start with the same phoneme. Similarly, the syntaxes are themselves likely to be significant here. In the (overly) simple syntactic formulation here, some words sit at the branch points between different syntactic trajectories, while some are unambiguously followed by other words as they only appear on one trajectory. Even at this very simple level, one could examine the influence of word predictability on both fluency and comprehension. However, a more complete account would likely need to take account of rapid plastic changes that allow for a more flexible expression of syntax (Sun & Manohar, 2023).

In Section 3, we alluded to an alternative formulation of the orbit dynamics. To recap, we have primarily reported on the impact of orbit dynamics in which all words start at position 1 but have variable end positions before transitioning to the space between words in position 8. The alternative formulation uses a variable start position, but all words end on position 7. This frontloads the uncertainty associated with each word, as any uncertainty about the word engenders uncertainty about the likely orbit state associated with the first phoneme. Perhaps unsurprisingly, this leads to greater sensitivity to the manipulations reported in Figure 4, with uncertainty about which word to say, and whether to say it, leading to longer gaps and greater phonemic repetition. A prediction we could derive from the different formulations is that, if our brains represent common start timings for words, speech dysfluency may be alleviated by external cueing timed with the start of words as this confirms the orbit position. In contrast, if our brains represent common endings, external cueing for the last phoneme may be more helpful.

One way in which the generative model itself could be extended is through inclusion of other (slower) levels of the temporal hierarchy that generates trajectories of words and the behaviour of conversational participants (Friston et al., 2017c). This would allow one to capture higher order statistics—for instance, those of switches between different word sequences. Crucially, explicit modelling of a hierarchy means that many of the parameters we have been concerned with at this level (effectively, as independent variables that can be manipulated to generate dependent—behavioural—variables) would be determined by what happens at a higher level. As an example, how might higher level beliefs we implicitly hold about the kinds of conversational partners we might encounter affect the parameters of Table 3? If one believes they are speaking to a relatively obnoxious conversational partner, they might predict that this person is likely to both interrupt them (high $b_5$), and to continue speaking regardless of whether they start speaking themselves (low $b_6$). In contrast, a more respectful conversant may show the opposite pattern. This means that, while the parameters of Table 3 can be specified and perturbed independently when one considers only a single hierarchical level, it is likely that they will covary and may be reducible to a smaller number of independent dimensions. This reinforces the need to consider the ways in which parameters interact (e.g., the analysis of Figure 7) in addition to understanding their individual effects.

## 8—Limitations

The analysis presented here is far from exhaustive. For instance, while we have illustrated predictions one might make based upon parameter interactions in Figure 7, there are many other interactions one could assess and measures one might use to score the speech patterns generated to assess the impact of those interactions. One could formalise this using the same scores applied to real speech, including the Stuttering Severity Instrument (Riley Glyndon, 1972), Apraxia of Speech Rating Scale (Strand et al., 2014), and elements of the Western Assessment Battery (Shewan & Kertesz, 1980).

The biggest limitation of this work is that it implicitly divorces speech from semantics. This is clearly a severe simplification that may need to be rectified in subsequent iterations of this modelling. It is striking that, even without explicit semantic elements, a rich variety of healthy and pathological speech behaviours can be simulated. Arguably, this approach could be licensed by the influential dual streams model of language (Hickok & Poeppel, 2004) in which there is a dissociation between dorsal regions more involved in the relationship between sounds and their articulation—our focus—and more ventral regions involved in the relationship between sounds and their meanings.

However, this does have an impact on the ability of the minimal model outlined here to capture some aspects of speech dysfluency. For instance, in people who stutter, the propensity to stutter for a given word is not entirely divorced from its semantic content (Brundage & Bernstein Ratner, 2022; Quarrington, 1965). Words with high informational content, that change the semantics of a sentence, such as the *X* in 'my name is *X*', can be associated with greater impairments in speech fluency. As outlined above, such points in a sentence, with multiple possible answers (which are also sensitive to who is speaking), might reflect branch-points in a linguistic sequence which are inherently more uncertain. Capturing communicative intent behind these words requires incorporation of a more semantically meaningful language model.

The visual element of this model represents a binary visual outcome—presence or absence of a conversational partner—only. While this was kept deliberately simple, to focus upon questions of private versus social speech, it clearly omits visual clues that are very important for verbal discourse. This includes things like lip-reading, facial expressions, or even sign-language. One could extend this by expanding the number of visual outcomes possible, replacing ['present', 'absent'] with ['lips moving', 'lips not moving', 'absent'], or even decomposing visible lip movements into phoneme-specific movements. Again, for our current purposes, the presence of another is designed to introduce ambiguity (i.e., one must do additional work to understand who is speaking), but clearly for speech comprehension visual clues play the opposite role, reducing ambiguity about what is being said. How might this change the results presented in this paper? The most obvious would be the results in Figure 6, showing something akin to receptive aphasia. In the presence of visual cues that are predictable based upon orbit position and the word being spoken, one might expect to be more resistant to attenuations in the precision of auditory phoneme prediction.

Finally, the validity of this modelling approach is only partially established by the results reported here. More specifically, there is a degree of face validity demonstrated in the ability to produce patterns of fluent speech-like behaviours, and auditory segmentation of sequences of speech, using the same underlying model. Of note, both active and perceptual elements of this rely upon sequences of phonemes of varying lengths as in real speech, supporting a set of phonemes comparable to those used in spoken English, and dealing with multimodal sensory data. This is further supported by prior work demonstrating the neurobiological plausibility of Active Inference as applied to POMDP models [e.g., (Friston et al., 2017a; Parr et al., 2019)]. The construct validity of our approach is illustrated by the realistic patterns of speech breakdown—capturing aspects of a range of patterns seen in human speech dysfluency syndromes—that were obtained through perturbations to model parameters. This specifically includes the impact of private versus public speech as seen in people who stutter (see Figure 7 and surrounding discussion).

The predictive validity has yet to be established. There are several ways one could approach this. First, as one can see from Figures 4-7, we can predict (given specific parameter perturbations), changes in both speech production and perception. This means we would predict co-occurrence of these findings when they occur, and a dissociation between production and perception would go against these predictions. Second, in progressive dysfluency syndromes (i.e., progressive non-fluent aphasia), one might predict that the relevant parameters that explain their syndrome will become progressively and monotonically perturbed with time. One could estimate these parameters at varying timepoints from recorded speech to establish whether this is the case. Third, the generative model implies very specific dependencies between variables, which implies specific patterns of sparse (e.g., synaptic) communication when inverting this model. In other words, we can make specific predictions about network architecture, summarised in Figure 8, and simulate timeseries signals from each node in this network. Such signals could be used as parametric regressors in functional imaging, to assess for the presence of these same signals during speech and to pin down their anatomical foci.

## 8—Conclusions

This paper is primarily designed to introduce a model of speech fluency, articulated in terms of Active Inference, as a starting point for further research looking at breakdown of fluency in people who stutter or those with nonfluent dysphasia syndromes. However, we can comment on a few things learned even from these initial simulations. Under the hypothesis that our brains use a common representation for the words spoken by ourselves and others, and that this comprises a variable length orbit and word representation:

- When one is sufficiently confident about whether they are speaking, and if so, what they wish to say, the same machinery can be used to generate speech and to segment words from auditory streams of phonemes while listening to another.
- Loss of precision in the word sequence results in a pattern of whole-word and start-of-word repetitions and pauses between words that are worsened in the presence of a listener. These features are typical of stuttering. Loss of precision in whether to speak is associated with phonemic omissions in addition to increased pauses between words.
- Loss of precision in sequencing of amongst different phonemes impairs initiation of speech with relative preservation of speech perception when listening. Conversely, loss of precision in predicting auditory phonemes does not speech generation as it can be compensated for by precise proprioceptive predictions.
- High prior expectations of being interrupted in the presence of another potential speaker is not a plausible explanation, on its own, for dysfluency syndromes, but may interact with the elements outlined above in causing stutter.

The novelty of these simulations is that they integrate speech output and perception with social cues. Consequently, they make specific predictions about how predictability of language interacts with social expectations and precision—which may be crucial in understanding disorders of speech.

## Software availability

The numerical simulations presented in this paper can be reproduced and customised using MATLAB code (tested in MATLAB 2025a) available from https://github.com/tejparr/Computational-Neurology. Specifically, the DEMO_Speech_Fluency.m routine will generate animations and belief-update plots from which our figures were generated.

**Acknowledgements**

TP is supported by an NIHR Academic Clinical Fellowship (ref: ACF-2023–13–013). SGM is supported by the National Institute of Health and Care Research (NIHR) Oxford Biomedical Research Centre (BRC), NIHR Oxford Health BRC, and MRC Clinician Scientist Fellowship (MR/P00878/X). BD was supported by the Dominic Barker Trust (Registered Charity no. 1063491). We are grateful to Kate Watkins for helpful comments on a draft of this manuscript.

**Declarations of interest**

The authors declare no known conflicts of interest.